\documentclass[twoside,reqno,12pt]{amsart}
 \usepackage[dvips,final]{graphicx} 
 \usepackage{hyperref}

 \newtheorem{thm}{\bf Theorem}[section]

 \newtheorem{lemma}[thm]{\bf Lemma}
 \newtheorem{corollary}[thm]{\bf Corollary}
 
 \newtheorem{conjecture}[thm]{\bf Conjecture}

 \numberwithin{equation}{section}

 \usepackage{latexsym}

 \usepackage{amsfonts,amssymb}

 \usepackage{enumerate}

 \def\vmiobmp#1#2#3#4#5#6#7{
  \noindent
  \vbox{
  \bigskip
  \vspace*{#2}
  \centerline{\hbox to #1{\special{bmp:#4.bmp x=#1 y=#2}}}
  \smallskip
  \didascalia{#6}
  \vskip -4pt
  \didascalia{#7}
 }}

 \def\Xmiobmp#1#2#3#4#5#6#7{
  \noindent
 \begin{figure}[!ht]
  \vspace*{#2}
  \centerline{\hbox to #1{\special{bmp:#4.bmp x=#1 y=#2}}}
  \caption{{#6} {#7}}
 \label{#4}
 \end{figure}
 }

 \def\nmiobmp#1#2#3#4#5#6#7{
  \noindent
 \begin{figure}[!h]
  \makebox[#2]{\framebox[#1]{\rule{0pt}{#2}}
  \centerline{\hbox to #1{\special{bmp:#4.bmp x=#1 y=#2}}}
  }
  \caption{#6}

  \end{figure}
 }

  \let\b=\beta   \let\d=\delta  \let\e=\varepsilon
  \let\g=\gamma       \let\l=\lambda
       \let\o=\omega      
   \let\s=\sigma \let\t=\tau   
   
 \let\D=\Delta   \let\G=\Gamma  \let\L=\Lambda 
 \let\O=\Omega

  \def\cB{{\cal B}} \def\cC{{\cal C}} 
   \def\cG{{\cal G}} \def\cH{{\cal H}}
    
 \def\cM{{\cal M}} \def\cN{{\cal N}}  
  \def\cR{{\cal R}} \def\cS{{\cal S}} 
    
  \def\cZ{{\cal Z}}

 \def\ldue{\ell_2}
 \def\ltre{\ell_3}
 
 \def\P{{\Bbb P}}
 \def\Q{{\Bbb Q}}
 \def\R{{\Bbb R}}
 \def\E{{\Bbb E}}
 \def\N{{\Bbb N}}
 \def\wt{\widetilde}
 \def\ddim{$d$-dimensional }
 \def\dmdim{$(d-1)$-dimensional }
 \def\err{(1+o(e^{-\b\d}))}
 
 \def\errhb{(1+o(h)+o(e^{-\b \d}))}
 
 \def\sselle{Z}
 \def\tonda#1{\left( #1 \right)}
 \def\quadra#1{\left[ #1 \right]}
 \def\graffa#1{\left\{ #1 \right\}}
 \def\TT#1#2{
     \P \tonda{\t^{#1}_{#2} < \t^{#1}_{#1}}}
 \def\Tzyx#1#2#3#4{
     \widetilde \P_{#4} \tonda{\t^{#1}_{#2} < \t^{#1}_{#3}}}
 \def\TTT#1#2#3{\Tzyx{#1}{#2}{#1}{#3}}
 \def\px{\check p}
 \def\py{\hat p}
 \def\pref#1{\cC_{#1}}
 \def\expd{e^{-\b \d}}
 \def\int#1{\lceil #1 \rceil}
 
 \def\plus{\mathbf {+1}}
 \def\minus{\mathbf {-1}}

 \def\acapo{\par\noindent}
 \def\Bbb{\mathbb}
 \def\cal{\mathcal}

 \def\Z{\Bbb Z}

\begin{document}
 \title [Metastability Beyond Exponential Asymptotics]
 {{Metastability in Glauber Dynamics in the Low-Temperature limit:
 Beyond Exponential
 Asymptotics}}

 \author[A. Bovier]{ Anton Bovier}
\address{Weierstrass-Institut f\"ur
 Angewandte Analysis und Stochastik,
Mohrenstrasse 39, 10117 Berlin, Germany}
\email{ bovier@wias-berlin.de}

 \author[F. Manzo]{ Francesco 
Manzo}
\address{Institut f\"ur Mathematik, TU-Berlin, Strasse des 17. Juni 
136, 10623 Berlin, Germany}
\address{ Adress from Aug. 1, 2001: Mathematisches Institut, 
Universit\"at Potsdam, PF 601553, Germany}
\email{ manzo@ mat.uniroma2.it }
\thanks{Supported by the DFG through the Graduiertenkolleg ``Stochastische
Prozesse und Probabilistische Analysis''}

\maketitle

 \centerline{\today}
 \begin{abstract}
 We consider Glauber dynamics of classical spin systems of Ising type
 in the limit when the temperature tends to zero in finite volume. We
 show that information on the structure of the most profound minima and
 the connecting saddle points of the Hamiltonian can be translated into
 {\it sharp} estimates on the distribution of the times of {\it
 metastable } transitions between such minima as well as the low lying
 spectrum of the generator. In contrast with earlier results on
 such problems, where only the asymptotics of the exponential rates is
 obtained, we compute the precise pre-factors up to multiplicative
 errors that tend to 1 as $T\downarrow 0$. As an example we treat the
 nearest neighbor Ising model on the 2 and 3 dimensional square
 lattice. Our results improve considerably earlier estimates obtained
 by Neves-Schonmann [NS] and 
Ben Arous-Cerf [BC] and Alonso-Cerf [AC]. Our results employ the methods
 introduced by Bovier, Eckhoff, Gayrard, and Klein in [BEGK1,BEGK2].
 \end{abstract}

\section{ \label{Section 1.} Introduction. } \
 Controlling the transitions from metastable states to equilibrium in
 the stochastic dynamics of lattice spin systems at low temperatures
 has been and still is a subject of considerable interest in
  statistical mechanics. The first mathematically rigorous results can
 be traced back to the work of Cassandro et al. [CGOV] that initiated
 the so-called ``path-wise approach'' to metastability. For a good
 review of the earlier literature, see in particular [Va]. All the
 mathematical investigations in the subject require some 'small
 parameter' that effectively makes the timescales for metastable
 phenomena 'large'. The somewhat simplest of these limiting situations
 is the case when a system in a finite volume $\L\subset \Z^d$ is
 studied for small values of the temperature $T=1/\b$. In systems with
 discrete spin space one is then in the situation where the dynamics
 can be considered as a small perturbation of a deterministic process,
 a situation very similar to what Freidlin and Wentzell  [FW] called
 'Markov chains with exponentially small transition
 probabilities'. Consequently, most of the work concerning this
 situation [OS1,OS2,CC,BC,N,NS] can be seen as extensions and
 improvements of the {\it large deviation }  approach initiated by
 Freidlin and Wentzell. This consists essentially in identifying the
 most likely path (in the sense of a sequence of transitions) and
 proving a large deviation principle on path-space. While this approach
 establishes very detailed information on e.g. the typical exit paths
 from metastable states, the use of large deviations methods entails a
 rather limited precision. Results for e.g. exit times $\t$ are
 therefore typically of the following type:
 For any $\e>0$,
 $$
 \P\left(e^{\b (\D-\e)} <\t<\e^{\b(\D+\e)}\right)\uparrow 1,\text{as
 $\b\uparrow\infty$}
 $$
 where $\D$ can be computed explicitly. Similarly, one has results on
 eigenvalues
 of the generator that are of the form
 $$
 \lim_{\b\uparrow\infty}\b^{-1}\ln \l_i(\b) =\g_i
 $$
 with explicit expressions for the $\g_i$ (see e.g. [FW,S]). From many
 points of view, the precision of such results is not satisfactory, and
 rather than just exponential rates, one would in many situations like
 to have precise expressions that also provide the precise
 pre-factors. This is particularly important if one wants to understand
 the dynamics of systems with a very complex structure of metastable
 states, and in particular {\it disordered } systems. (For a rather
 dramatic illustration, see e.g. [BBG1,BBG2] where aging phenomena in the
 random energy model are studied.) Another drawback of the large
 deviation methods employed is that they are rather heavy handed and
 require a {\it very detailed knowledge } of the entire energy
 landscape, a requirement that  frequently cannot be met.

 In two recent papers [BEGK1,BEGK2] a somewhat new approach to the
 problem of metastability has been initiated aiming at improving the
 precision of the results while reducing at the same time the amount of
 information necessary to analyze
 a given model. To achieve this goal,
 the attempt to construct the precise exit paths is largely abandoned,
 as are, to a very large extent, large deviation methods.

 The general structure of this approach is as follows. In [BEGK2]
 the notion of a {\it set of metastable points } is introduced. The
 definition of this set employs only one type of objects, namely
 {\it Newtonian capacities} (which may also be interpreted as {\it
 escape probabilities}). If such a set of metastable points can be
 identified, [BEGK2] provides a general theorem that yields precise
 asymptotic formula for the {\it mean  exit time} from each
 metastable state, shows that this time is asymptotically
 exponentially distributed (in a strong sense), and
 states that each mean exit time is the inverse of one small
 eigenvalue of the generator. While [BEGK2] assumes reversibility
 of the dynamics, in [E] is shown that almost the same results can
 be obtained in the general case. Thus, the analysis of
 metastability is essentially reduced to the computation of
 Newtonian capacities. The great advantage of such a result is that
 capacities are particularly easy to estimate, due to the fact that
 they verify a particularly manageable {\it variational principle}.
 This fact is well known and has been exploited in the analysis of
 transience versus recurrence properties of Markov chains (see e.g.
 [DS]); however, its particular usefulness in the context of
 metastability seems to have been noticed only in [BEGK1] where it
 was used in the context of reversible discrete diffusion processes
 motivated from certain mean field spin systems.

 In this short paper we will show that the approach is even more
 efficient and simple in the context of the zero temperature limit
 of Glauber dynamics of spin systems in finite volume. We will show
 that in rather general situations, capacities in this limit can be
 computed virtually exactly in terms of properties of the {\it
 energy landscape}, and therefore all interesting properties of the
 dynamics can be inferred from a (not overly detailed) analysis of
 the energy landscape generated by the Hamiltonian considered. As a
 particular application that should illustrate the power of our
 approach, we apply the general results to the Ising model (in two
 and three dimension).

 \section{ \label{Section 2.} The general setting and the main theorem. } \

 In this section we set up the general context to which our results
 will apply. It will be obvious that Glauber dynamics of finite volume
 spin systems at low temperatures provide
 particular examples. We will consider Markov processes on a finite
 state space $\O$ (the {\it configuration space}). To define the
 dynamics, we need the following further objects.

 \begin{enumerate}
 \item  A {\it connected graph}  $\cG$ on $\O$.
 We denote by $E(\cG)$ the set of the edges in
 $\cG$.
 \item A {\it Hamiltonian} $H : \O \to \R$ also called {\it energy}.
 \item The {\it Gibbs measure}
   $\Q(x) := \frac{1}{\cZ} \exp(- \b H(x))$, where $\cZ$ is the
   normalization factor called partition function, and $\b$ is the {\it
 inverse temperature.}
 \end{enumerate}
 We consider {\it transition probabilities } $P(x,y)$ such that
   if $\graffa{x,y}\in E(\cG)$, $P(x,y)>0$, and $P(x,y)=0$ if
 $x\neq y$ and $\graffa{x,y}\not\in E(\cG)$. We assume moreover that
   the transition probabilities are  reversible with respect to the Gibbs
 measure, i.e.
   \begin{equation}
     \label{eq:bilanciodettagliato}
     \Q(x) P(x,y) = \Q(y) P(y,x) .
   \end{equation}
 We will also make the simplifying assumption that any existing
   transition in the graph is reasonably strong, i.e. we assume that
   there exists a constant $C>0$ such that\footnote{The
   constant $C$ will typically be of the order of the inverse of
   the maximum coordination number of the graph $\cG$.}
  \begin{equation}
    \label{eq:haste}
     P(x,y) + P(y,x) \geq C \ \ \forall \; \graffa{x,y} \in E(\cG),
   \end{equation}
 by reversibility, \eqref{eq:haste} is equivalent to
  \begin{equation}
    \label{eq:haste2}
     P(x,y)  \geq \frac{C}{1+\exp (-\b(H(y)-H(x)))} \ \
     \forall \; \graffa{x,y} \in E(\cG).
   \end{equation}

 To be able to state our results we need some further notations.
 \begin{enumerate}
 \item Given a one-dimensional subgraph $\o$, we write $\o: x \to I$
   if the subgraph has one end in $x$ and the other end in $I$.
   One dimensional subgraphs have a natural parameterization
   $\o_0,\dots,\o_K$, where $K:=|\o|-1$,
   $\forall k=0,\dots , K-1 \ $ $q(\o_k,\o_{k+1})>0$ and
   $\o: \o_0 \to \o_K$.
 \item Let $\widetilde H (\{ \o \}) := \max_{z \in \o} H(z)$.
   For $x\in \O$ and $I\subset \O$, we introduce the
  {\it communication height}, $\widehat H(x,I)$, between $x$ and $I$ as
    \begin{equation}
 \label{com1} \widehat H (x,I) := \min_{\o : x \to I} \widetilde
 H(\{\o\}).
 \end{equation}

 Moreover we define the set of {\it saddle points} for $x$ and $I$ by
   \begin{equation}
    \label{eq:selle}
     \cS_{x,I} := \graffa{ z \in \O \; ; \; \exists \; \o : x \to I
     \text{ with } z \in \o \text{ and } H(z) = \widehat H(x,I) }
   \end{equation}
 \item Furthermore, we define the set of points
  \begin{equation}
 \label{com2}
 D_x^I := \graffa{z \; ; \; H(\cS_{z,x}) < H(\cS_{z,I})}
 \end{equation}
 These will be the points that are 'closer' to $x$ than to $I$.
 \item For any set $A\subset \O$, we define its {\it outer
 boundary} $\partial A $ as the set of all points in $A$ from which an edge
 of $\cG$  leads to its complement, $A^c$.
 \item For $z \in \partial D_x^I$, let
   $\px_z:= \sum_{y' \in D_x^I} P (z,y')$  and
   $\py_z:= \sum_{x' \in \O \setminus D_x^I} P (z,x')$.
   We let
   \begin{equation}
    \label{pre-factor}
     \pref{x,I}:=  \sum_{z \in \cS_{x,I}} \frac{\py_z \px_z}{\py_z + \px_z}.
   \end{equation}

 \item Let $W_x := \graffa {y \; ; \; H(y) < H(x)} $.
   For $x \in \O$, we set
   $\G(x) := H(\cS_{x,W_x}) - H(x)$.
   If $x$ is not a local minimum of the Hamiltonian, $\G(x)=0$.
   If $x$ is a global minimum of the Hamiltonian, we set $\G(x)=\infty$.
 \item For the process $X_t$ starting at $x$, we define the
   {\it hitting time} to the set $I \subset \O$ as
   $\t^x_I:=\inf \{0<t \in \N \; ; \; X_t \in I \}$.
 \item We denote by $\wt \cM$ the set of all {\it local minima} of
 $H$. We call a subset $\cM\subset \wt\cM$ a set of {\it metastable
 states}, if a point that realizes the absolute minimum of $H$ is
 contained in $\cM$ and if, for all $y\in \wt\cM\backslash\cM$,
 $\G(y)<\min_{x\in\cM}\G(x)$.
 It is important to realize that for given $H$, $\cM$ may often be
  chosen in different ways.
 The idea will be that we will observe the process only at its
  visits to $\cM$.
 Thus, the actual choice of $\cM$ will depend on how much
  information we want to retain about the detailed behavior of the
  process.
 Note that this definition implies
 that for all $z\not\in\cM$, $\widehat H(z,\cM)\geq \widehat
 H(z,W_z)$.
 \item Finally, for $x \in \cM$, we need to define the quantity
 $$
 N_x:= \#\graffa{z\in \O: \big\{H(z)=H(x)\big\} \cap
 \big\{\widehat H(x,z) < \widehat H(x,\cM\backslash x)\big\}},
 $$
 which represents the degeneracy of the minima of the Hamiltonian.

 \end{enumerate}

 We can now formulate the main general result of this paper in the
 general setting. Let us consider some set $\cM$ of metastable
 points. To be able to formulate concise and general results, we
 make some further assumptions that will be true for ``generic''
 Hamiltonians.\footnote{Note that for any Hamiltonian, one may
 select different sets of metastable points. The requirements (h1)
 and (h2) depend on the Hamiltonian as well as on the choice of
 $\cM$. }

 \begin{enumerate}
 \item [(h1)] For any $x\neq y\in\cM$, $\G(x)\neq \G(y)$.
 \item [(h2)]For any $x\neq y\in\cM$, $\cS_{x,y}$ consists of isolated single
 points\footnote{
 In the appendix we will explain how one can proceed to obtain comparable
 results in the case when this condition is not satisfied.}.

 \end{enumerate}

 \begin{thm}
 \label{th1}
 Let $\cM$ be a set of metastable states for the Hamiltonian $H$ satisfying the
 conditions (h1) and (h2) above. For $x\in \cM$, set
 $ \cM_x :=\graffa{y\in \cM \; ; \; H(y)<H(x)}= \cM \cap W_x$.
 Let $ \t(x) := \t^x_{ \cM_x}$,
 $\cS := \cS_{x,W_x}$ and
 $\pref{x} := \pref{x,W_x}$.\
 Then there exists $\d>0$, independent of $\b$, such that for 
 any $x\in\cM$,
 \begin{enumerate}
 \item [(i)]
       \begin{equation}
	 \label{eq:th31i}
	 \E \; \t(x)
	  =
	 N_x\pref{x}^{-1} e^{\b \G(x)} \err
       \end{equation}
 \item [(ii)] there exists an eigenvalue
       $\l_x$ of $1-P$ such that
       \begin{equation}
	 \label{eq:th31ii}
	 \l_x= \frac 1{\E  \; \t(x)}\err
       \end{equation}
 \item [(iii)] if $\phi_{x}$ is the right-eigenvector of $P$
       corresponding to $\l_x$, normalized so that
       $\phi_x(x)=1$, then
       \begin{equation}
	 \label{eq:th31iii}
	 \phi_x(y)=
	 \P \tonda{ \t^{y}_{x} <
	      \t^{y}_{\widetilde \cM_x}} + o(e^{-\d\b}))
       \end{equation}
 \item [(iv)]
       \begin{equation}
	\label{eq:th31iv}
	 \P \tonda{ \t(x) > t \; \E \; \t(x)}
	 =
	 e^{- t \err} \err
       \end{equation}
 \end{enumerate}
 \end{thm}

 In Section \ref{Section 6.}, we will apply Theorem 2.1 to a well
 known situation, the kinetic Ising model,
 in the limit of
 vanishing temperature.

 Let us anticipate our main result about the kinetic Ising model,
 referring to Section \ref{Section 6.} for precise definitions and
 notation.
 \begin{thm}\label{mainising1}
 Consider the kinetic
 Ising model with Metropolis dynamics  in
 dimension $d=2$ or $d=3$ in a torus $\L^d(l)$ with diameter $l$.
 The magnetic field $0<h<1$ is chosen such that $2(d-1)/h$ is not
 an integer.
 Then, \acapo
 the two configurations $\minus$ (all minus spins) and
   $\plus$ (all plus spins) form a metastable set (if the magnetic
   field $h$ is positive, $\plus$ is stable), and
 \begin{itemize}
   \item In dimension $2$, let $\ldue:=\lceil \frac{2}{h}\rceil$
   and $\G_2:= 4 \ldue -h \tonda{\ldue^2 - \ldue + 1}$ be the
   diameter and the activation energy of the "critical droplet",
   respectively.
   Then,
   \begin{equation}\label{misi2}
     \E \t(\minus) = \frac{3}{8} \frac{1}{\ldue-1} e^{\b \G_2} \err
     =\frac{3}{16} h e^{\b \G_2} \errhb
   \end{equation}
   \item In dimension $3$, let $\ltre:=\lceil \frac{4}{h}\rceil$
   and
   $$
\G_3:= \tonda{6 \ltre^2 -4 \ltre +4 \ldue} -
   h \tonda{\ltre^3 - \ltre^2 + \ldue^2 - \ldue + 1}
$$ be the
   diameter and the activation energy of the "critical droplet",
   respectively.
   Then,
   \begin{equation}\label{misi3}
     \E \t(\minus) =
       \frac{1}{16} \frac{1}{(\ltre - \ldue +1 )(\ldue-1)}
       e^{\b \G_3} \err
     = \frac{1}{256} h^2 e^{\b \G_3} \errhb
   \end{equation}
 \end{itemize}
Here as in Theorem 2.1, $\d>0$ is independent of $\b$ (but depends on 
arithmetic properties of $h$). 
\end{thm}

 {\bf Remark.} Note that in our model we flip at most one spin per time step.
 In continuous time dynamics the mean transition times would be lowered by a 
 factor $1/|\L|$.

 The above Theorem shows how the results of Theorem \ref{th1} can
 be applied (via the analysis of the energy landscape carried out
 for the Ising model in [NS] and [AC,BC]) to the so-called
 {Freidlin-Wentzell} regime. Notice that the methods of [BEGK2] can
 be applied in a very similar way to situations where the volume
 grows with $\b$ to compute "exactly" the probability of first
 appearance of a critical droplet (a preliminary problem for the
 infinite-volume metastability carried out in [DeSc]).

 \section{ \label{Section 4.} Basic tools. } \

 Theorem \ref{th1} relies on  Theorem
 1.3 in [BEGK2] that links relative  capacities of metastable sets to mean
 exit times and to the low lying spectrum of $1-P$. The additional work needed to
 prove Theorem \ref{th1} will be to estimate capacities in terms of the
 Hamiltonian $H$, and
  to show that the hypotheses of Theorem
 1.3 in [BEGK2] are satisfied in our setting.

 Let us state Theorem
 1.3 in [BEGK2]  specialized to  our case.

 In their context, a set $\cM \in \O$ is called a set of
 {\it metastable points} in the sense of [BEGK2] if
 \begin{equation}
  \label{eq:defmeta}
   \frac{
     \sup_{{x\not =y \in \cM}} \TT{x}{y} }
    {\inf_{z \in \O}  \P \tonda{\t^{z}_{\cM} \le \t^{z}_{z}} }
   \to 0
   \text { as } \b \to \infty.
 \end{equation}
 The set $\cM$ is {\it generic} in the sense of [BEGK2] if
 for any $x,y\in\cM$, and $I\subset \cM$,
 $\frac {\P\left(\t^x_I<\t^x_x\right)}{
 \P\left(\t^y_I<\t^y_y\right)}$ tends either to zero or to infinity, as
 $\b\uparrow\infty$, and if the absolute minimum of the Hamiltonian is
 not degenerate.

 \begin{thm}  {\rm (Theorem  1.3 in [BEGK2])}
 \label{th41}
 Let $\cM$ be a generic set of metastable states
 in the sense of [BEGK2],
 and let for $x\in\cM$,
  $\cM_x$ and $\t(x)$ be defined as in Theorem \ref{th1}.
 Then, for any $x\in\cM$, the following holds:
 \begin{enumerate}
 \item [(i)]
       \begin{equation}
	 \label{eq:th13i}
	 \E \; \t(x)
	  =
	 \frac{N_x}{\TT{x}{\cM_x}}(1+ o(1))
       \end{equation}
 \item [(ii)] for any $x \in \cM$, there exists an eigenvalue
       $\l_x$ of $1-P$ such that
       \begin{equation}
	 \label{eq:th13ii}
	 \l_x= \frac{1}{\E \; \t(x)}(1+ o(1)) ,
       \end{equation}
       moreover, the eigenvalues of $1-P$
       not corresponding to any $x \in \cM$
       are in the interval
       $(c |\O|^{-1} \inf_{z \in \O} \TT{z}{\cM},2]$
       for some positive constant $c$.
 \item [(iii)] if $\phi_{x}$ is the right-eigenvector of $P$
       corresponding to $\l_x$, normalized so that
       $\phi_x(x)=1$, then
       \begin{equation}
	 \label{eq:th13iii}
	 \phi_x(y)=
	 \P \tonda{\t^{y}_{x} < \t^{y}_{\cM_x}} + o(1)
       \end{equation}
 \item [(iv)] for any $x \in \cM$, for any $t>0$,
       \begin{equation}
	\label{eq:th13iv}
	 \P \tonda{\t(x) > t \; \E\; \t(x)}
	 =
	 e^{- t (1+ o(1))} (1+ o(1)).
       \end{equation}
 \end{enumerate}
Here $o(1)$ stands for a small error that depends only on the small parameters
introduced via \eqref{eq:defmeta} and the non-degeneracy condition following it. 
\end{thm}

 We leave it to the reader to verify that this theorem is indeed a special
 case of the more general result stated in [BEGK2].

 Theorem \ref{th1} will follow from Theorem \ref{th41}
 since  in the finite-volume and $\b \to \infty$ regime, we
 compute $\TT{x}{\cM_x}$ and show that local minima of the
 Hamiltonian are metastable states giving at the same time the
 value of the nucleation rate in the limit $\b \to \infty$.

 The key estimate is the following Lemma.
 \begin{lemma}
 \label{lemma1}
 $\forall x,y \in \wt\cM$
 such that $\cS_{x,y}$ is a set of isolated single points,
 \begin{equation}
   \label{eq:lemma1}
   \TT{x}{y}
   =
   \pref{x,y} e^{-\b (H(\cS_{x,y}) - H(x))} (1+o(\expd)).
 \end{equation}
 \end{lemma}
 We will explain in the appendix how our method can be extended to
 situations where the saddles are degenerate. In this case the
 pre-factor $\pref{x,y}$ does not have the nice form in
 \eqref{pre-factor} but
 can still be computed explicitly in terms of small
 "local variational problems".

 \begin{lemma} \label{cor1} Let $x$ be a minimum for the
 Hamiltonian. Then, $x$ is a metastable state (in the sense of
 [BEGK2]) in the set $\cM:=\graffa{y \; ; \; \G (y) \geq \G(x) }$.
 \end{lemma}

 Clearly,  Theorem \ref{th1} immediately follows from Theorem \ref{th41},
  Lemma \ref{lemma1} and Lemma \ref{cor1}.

 \section{ \label{Section 5.} Proof of Lemmata
	    \ref{lemma1} and \ref{cor1}. } \

 In order to prove Lemmata \ref{lemma1} and \ref{cor1},
 we make use of many ideas
 contained in [BEGK1].

 The following Lemma corresponds to Theorem 6.1 in [Li].
 \begin{lemma}
 \label{lemma2}
 (Dirichlet representation).\acapo
 Let $\cH^x_y := \graffa{h: \O \to [0,1] \; ; \; h(x)=0, h(y)=1}$ and
 \begin{equation}
   \label{eq:dirichlet}
   \Phi(h):=  \frac{1}{\cZ}  \sum_{x',x'' \in \O}
    e^{-\b H(x')} P(x',x'') [h(x')-h(x'')]^2.
 \end{equation}
 Then,
 \begin{equation}
  \label{eq:lemma2}
    \frac{ e^{-\b H(x)}}{\cZ } \TT{x}{y}
   =
  \frac{1}{2}\inf_{h \in \cH^x_y} \Phi(h)
 \end{equation}
 \end{lemma}
 \begin{proof}
 See [Li], Chapter II.6.
 \end{proof}

 Note that the left-hand side of \eqref{eq:lemma2}
 has the potential-theoretic
 interpretation of the {\it Newtonian capacity} of the point $y$ relative to
 $x$ (i.e. the electric charge induced on the  grounded site $x$
 when the potential is set to 1 on the site $y$).
 The Dirichlet form is just the
 electric energy, and the minimizer $h^*$ is the {\it equilibrium
 potential}, with the probabilistic interpretation
 $h^*(z)=\P\left(\t^z_y<\t^z_x\right)$.

 The strength of this variational representation comes from the monotonicity
 of the Dirichlet form in the variables $P(x',x'')$,
 expressed in the next Lemma, known as Rayleigh's short-cut rule
 (see Lemma 2.2 in [BEGK1]):

 \begin{lemma}
 \label {lemma3}
 Let $\D$ be a subgraph of $\cG$
 and let $\widetilde \P_\D $ denote the law of
 the Markov chain with transition rates, for $u \not = v$,
 defined by
 $\widetilde P_\D (u,v) := P(u,v) \Bbb I \graffa{ \graffa{u,v} \in E(\D)}$.
 If $x$ and $y$ are vertices in $\D$, then
 \begin{equation}
  \label{eq:short}
   \TT{x}{y} \ge \TTT{x}{y}{\D}
 \end{equation}
 \end {lemma}
 \begin{proof}
 The proof follows directly from Lemma \ref{lemma2}
 and can be found in  [BEGK1].
 \end{proof}

 The following Lemma corresponds to Lemma 2.3 in [BEGK1] and is a well known
 fact (see e.g. [DS]).

 \begin{lemma}
 \label {lemma4} {\rm (The one dimensional case)}.

 Let $\o$ be a one-dimensional subgraph of $\cG$,
 $K:=|\o|-1$
 and let
 $\{\o_n\}_n:\{0,\dots,K\} \to \O$
 be such that
 $\forall n \le K$,
 $q(\o_n,\o_{n-1}) > 0$
 \begin{equation}
  \label{eq:dim1}
   \TTT{\o_0}{\o_{K}}{\o}
   =
   \quadra{ \sum_{n=0}^{K-1}
    \frac{e^{-\b(H(\o_0)-H(\o_n))}}
	 {P(\o_{n},\o_{n+1})}
   }^{-1}
 \end{equation}
 \end{lemma}

 Remark: Lemmata \ref{lemma3}, \ref{lemma4} and  \eqref{eq:haste}
 immediately give the following bound: $\forall x',I$ s.t.
 $\cS_{x',I}$ is made of simple points,
 \begin{eqnarray}
   \TT{x'}{I}
   & \ge &
   \quadra{
    \sum_{n=0}^{K-1}
    \frac{
     e^{-\b(H(x')-H(\o_n))} }
    {P \tonda{\o_{n},\o_{n+1} } }
   }^{-1}
   \ge
 \nonumber\\
   \label{eq:apriori1}
   & \ge &
     C   \quadra{
    \sum_{n=0}^{K-1}
     \tonda{e^{-\b(H(x')-H(\o_n))}
     + e^{- \b (H(x')- H(\o_{n+1}))} }
   }^{-1}
 \nonumber\\
  \label{eq:apriori}
   & \ge &
   \frac{C}{2} e^{- \b (H(\cS_{x',I})- H(x'))} \tonda{1-\expd}
 \end{eqnarray}
 for any choice of the subgraph $\o : x' \to I$
 having its maximum energy in $\cS_{x',I}$.
 The constant $C$ is the same as in \eqref{eq:haste2}.

 \begin{proof} [{Proof of Lemma \ref{lemma1}.}]

 Let $\G :=H(\cS_{x,y}) - H(x)$.

 We consider the surface $\sselle:= \partial D_x^y$.
 Notice that:
 \begin{enumerate}
   \item $\cS:= \cS_{x,y} \subset \sselle $
   \item $\exists \d > 0$ such that
     $\forall z \in \sselle \setminus \cS$,
     $H(z) \ge H(\cS) + \d$.
   \item  $\sselle$ is the outer boundary of a connected set
       that contains $x$.
 \end{enumerate}
 {\bf Remark:} In what follows, any other surface with properties 1, 2, and 3
 would give the bounds we need for the proof.
 The quantity $\cC'_{x,y}$ defined with respect to the new surface
 differs from $\pref{x,y}$ by a factor $1+o(\expd)$.

 We set
 $D_x := D_x^y$,
 $D_y := \O \setminus (\sselle \cup D_x)$,
 $\sselle^- := \partial \sselle \cap D_x$ and
 $\sselle^+ := \partial \sselle \cap D_y$.
 \acapo
 (1) The upper bound.

 We use Lemma \ref{lemma2} with
 $h(x'):=0$ if $x' \in D_x$ and
 $h(y'):=1$ if $y' \in D_y$;
 we choose $h(z)$ for $z \in \sselle$
 in an optimal way. In the rest of the space we choose
$h(x)=1$.

 We have
 \begin{equation*}
   \TT{x}{y}
   \le
   \frac{\cZ e^{\b H(x) }}{2 } \Phi(h)=
 \end{equation*}
 \begin{equation}
  \label{eq:low1}
   =
   \sum_{z \in \sselle}
   e^{-\b (H(z)-H(x))}
   P (z,x)
   \tonda{
   \px_z h^2(z) + \py_z (1-h(z))^2} +o(e^{-\b\d}),
 \end{equation}
 where we used reversibility. The small error comes from the mismatch 
on the boundary of $D_x$ that lies higher than the saddle hight. 

 The quadratic form
 $\px h^2 + \py  (1-h)^2$
 has a minimum for
 $h=\frac{\py }{\py  + \px}$.
 Hence, we can saturate the inequality
 \eqref{eq:low1} and get
 \begin{equation}
  \label{eq:low}
   \text{(l.h.s. of \eqref{eq:low1}) }
   \le
   \pref{x,y} e^{-\b \G} \tonda{1+\expd}.
 \end{equation}

 \acapo
 (2) The lower bound.

 We consider the subgraph $\D$ obtained by
 cutting all the connections to the vertices in
 $\sselle \setminus \cS$.

 We use Lemma \ref{lemma3} to bound the original process by the
 restricted process.

 We use \eqref{eq:apriori} to estimate the probability
 to reach $x' \in \sselle^-$
 and the probability to go from $y'\in \sselle^+$ to $y$: \acapo
 By the strong Markov property at time $\t^x_\cS$, we have
 \begin{eqnarray}
  \label{up1}
   \TTT{x}{y}{\D}
  & =&
   \sum_{z \in \cS}
     \widetilde \P_{\D} \tonda{\t^x_{z} \le \t^x_{\cS \cup x}}
     \widetilde \P_{\D} \tonda{\t^{z}_y < \t^{z}_{x}}
   \nonumber\\\label{upa}
 &  =&
   e^{-\b \G}
     \sum_{z \in \cS}
     \widetilde \P_{\D} \tonda{\t^z_{x} < \t^z_{\cS}}
     \widetilde \P_{\D} \tonda{\t^{z}_y < \t^{z}_{x}},
 \end{eqnarray}
 where we used reversibility.

 Now,
 \begin{equation}
  \label{eq:parte1}
   \widetilde \P_{\D} \tonda{\t^z_{x} < \t^z_{\cS}}
   =
   \sum_{x' \in \sselle^-}
     P \tonda{z,x'}
     \widetilde \P_{\D} \tonda{\t^{x'}_x < \t^{x'}_{\cS}}.
 \end{equation}

 We bound the last factor using a standard renewal argument (see
 e.g. [BEGK1] Corollary 1.6) that yields if $z' \in D_x$ the last
 term is exponentially close to $1$:
 \begin{eqnarray}
  \label{eq:b1}
   \Tzyx{x'}{\cS}{x}{\D}
  & =&
   \frac{ \Tzyx{x'}{\cS}{x \cup x'}{\D} }
	{ \Tzyx{x'}{x \cup \cS}{x'}{\D} }
   \le
   \frac{ e^{- \b \G}
	  \sum_{z \in \cS}
	  \Tzyx{z}{x'}{x \cup \cS}{\D} }
	{ \Tzyx{x'}{x}{x'}{\D} } \nonumber\\
  \label{eq:b2}
  &\le&
   \frac{ | \cS | e^{- \b \G} }
	{  C e^{-\b (\G - \d')} \tonda{1 -e^{-\b \d'}}}
   \le \expd,
 \end{eqnarray}
 where we used \eqref{eq:apriori}. By  putting together
 \eqref{eq:parte1} and \eqref{eq:b2} we get
 \begin{equation}
 \label{p1}
   \widetilde \P_{\D} \tonda{\t^z_{x} < \t^z_{\cS}}
   \ge
   \px_z \tonda{1-\expd}
 \end{equation}

 We use the same procedure to bound the last term in
 \eqref{upa}:
 \begin{equation}
  \label{eq:parte2}
   \Tzyx{z}{y}{x}{\D}
   \ge
   \sum_{y' \in \sselle^+}
     P \tonda{z,y'}
     \Tzyx{y'}{y}{x}{\D}.
 \end{equation}
 Again, the same arguments leading to \eqref{eq:b2}  show that
 the last term in this sum is exponentially close to 1:
 \begin{equation}
  \label{eq:c1}
   \Tzyx{y'}{x}{y}{\D}
   \le
   \frac{ | \cS | e^{- \b (H(z) - H(y')) } }
	{ e^{- \b (H(\cS_{y,y'}) - H(y')) }
	  \tonda{1 - e^{-\b\d'} } }
   \le
   \expd
 \end{equation}
 We put together \eqref{eq:parte2} and \eqref{eq:c1}
 and get
 \begin{equation}
  \label{p2}
   \widetilde \P_{\D} \tonda{\t^{z}_y < \t^{z}_{x}}
   \ge
   \frac{\py_z}{\py_z+\px_z} (1-\expd)
 \end {equation}
 Going back to \eqref{up1},
 we get from \eqref{p1}, \eqref{p2}
 \begin{equation}
  \label{up2}
  \widetilde \P_{\D} \tonda{\t^x_y < \t^x_x}
   \ge
   \pref{x,y} e^{- \b \G}    \tonda{1- \expd}
 \end {equation}
 \end {proof}
 \begin{proof}[Proof of Lemma \ref{cor1}]
 For any  $z \not \in \cM$
 we know that by definition of $\cM$, we have that
 $\G(z)<\min_{x\in\cM}\G(x)\equiv \G$. In view of Lemma \ref{lemma2} and
 the lower
 bound (4.5) we only need to show that this implies that
 $\widehat H(z,\cM)-H(z)<\G$.
 Now let $u\not\in\cM$ be the point that realizes the minimum of
 the energy among the states such that
 $\widehat{H}(z,u) < \widehat{H}(z,\cM)$.
 For such a point, by definition, $\widehat
 H(u,\cM)-H(u)=\G(u)<\G$. But  clearly,
 $\widehat{H}(z,\cM)-H(z) \le \widehat{H}(u,W_u)-H(u) < \G(x)$, and we are
 done.
 \end{proof}

 \section{ \label{Section 6.} The Ising case. } \

 In this section we want to illustrate the strength of Theorem
 \ref{th1} in a well known context, namely the stochastic Ising
 model on the $d$-dimensional lattice. In this case the state space
 is $\O=\{-1,+1\}^\L$, where $\L=\L(L)$ is a torus in
 $\Z^d$ with side-length $L$.
 For $\s \in \O$, the Hamiltonian is
 then given by
 \begin{equation}
  \label{isiham}
   H(\s)\equiv H_\L(\s)
   =-\frac{1}{2} \sum_{<i,j> \in\L}
   \s_i\s_j-h\sum_{i\in\L}\s_i,
 \end{equation}
 where the first sum concerns all the pairs of nearest neighbor
 sites in $\L$.

 Let $\s^i$ be the configuration that differs from $\s$ only in the
 value of the spin of site $i$ and $[a]_+$ denote the positive part
 of the real number $a$.
 We will consider for definiteness only the
 case of the Metropolis dynamics, i.e. the transition probabilities are
 chosen
 \begin{equation}
 \label{metro1} P(\s,\s')=\frac {e^{-\b[H(\s')-H(\s)]_+}}{|\L|},
 \,\hbox{ if }\, \s'=\s^i, i\in\L
 \end{equation}
 \begin{equation}
 \label{metro2}
 P(\s,\s)= 1-\sum_{i\in\L} P(\s,\s^i)
 \end{equation}
 and all others are zero.

 We will use the estimate given in
 Theorem \ref{th1} to analyze this dynamics in a finite volume
 $\L$, under a positive magnetic field, in the limit when
 $\b\uparrow\infty$. 

 Let $\minus$ and $\plus$ be the configurations  full of minuses
 or  full of pluses, respectively.
 We will show in Lemma \ref{pozzi} that $\graffa{\minus,\plus}$
 is a set metastable states.
 Apart from this characterization, we will only use the description
 of the energy landscape given 
 in [NS], [AC,BC] and [N] in dimension 2, 3 or larger, respectively.
 We will show that the methods of Theorem \ref{th1} allow to improve 
 the known estimates without requiring further analysis of the 
 energy landscape.
 In dimension 2 and 3, the improvement amounts to the computation
 of the exact (including the pre-factor $\pref{\minus,\plus}$) asymptotic
 value of the expected transition time $\t^-_+$
 needed to reach $\plus$
 starting from $\minus$
 (that by Theorem \ref{th41} is the inverse 
 of the spectral gap of $P$).
 In higher dimension, where our knowledge of the energy landscape is 
 not so detailed, we cannot compute the pre-factor but we show that 
 it is a constant independent of $\b$, while previous results 
 only gave sub-exponential bounds.

 We remark that, unlike the exponential factor $\exp (- \b \G(\minus))$
	  (that only depends on the
	  graph structure), the pre-factor
	  $\pref{\minus,\plus}$ is related to the particular  
	  Glauber dynamics we
	  choose.

 We consider $0<h<1$ and
 assume that $\forall k \le d-1$, $\frac{2 k}{h}$ is not an
 integer.

 We set $\ell_d := \lceil \frac{2 (d-1)}{h}\rceil$.

 A \ddim parallelepiped with all sides of length  $\ell-1$ or $\ell$
 is called {\it quasi-cube} in dimension $d$ with maximal
 side-length $\ell$.

 Given  a \ddim parallelepiped
  $(l_1 \times l_2 \ldots \times l_d)$ and a
  $(d-1)$-dimensional configuration
  $\eta^{d-1} \in \Z^{(l_2 \times l_3 \ldots \times l_d)}$,
  let us consider a configuration where the
  sites in the parallelepiped as well as  the sites of the form
  $l_1+1,i_2, \ldots, i_d$
  where $\eta^{d-1}(i_2,\ldots ,i_d)=+1$ have plus spin and
  all other sites have minus spins.
 For such a configuration, as well as for all its rotations and
  translations, we say that $\eta^{d-1}$ is
  {\it attached} to the parallelepiped.

 Following [N], we introduce a set  $\cB^d (v)$ in a recursive way:
  let $\cB^1(v)$ be the set of configurations where the pluses form  
  a slab with volume $v$,
  $\cB^d(v)$ is defined as the set of all configurations with volume $v$
  in the form of the  \ddim quasi-cube with maximal volume $v'\le v$
  with a \dmdim configuration $\eta \in \cB^{d-1}(v-v')$
  attached to one of its largest faces.
  Heuristically, these configurations are as close as possible to
  a cube.
 It is easy to see that the energy is constant in  every set
  $\cB^d(v)$; we will denote this energy by $H(\cB^d(v))$.

 We make use of the following Theorem
 \begin{thm}
 \label{neves} {\rm (Theorem 3 in [N])}\acapo
  In the whole \ddim lattice $\Z^d$,  $\cB^d(v)$ is a subset of the
the minimizer of the
  Hamiltonian in the manifold with  volume $v$.
 \end{thm}

 This result can be transported to the torus $\L(L)$ 
  only for sufficiently
  small values of $v/L^d$.
 For large values of $v$, the boundary conditions affect the shape
  of the minimizing configurations.
 Let
  $m:= \min \graffa{v \ge 1 ; H(\cB^d(v)) \le H(\minus)}$.
 We take $L$
  so large that the configurations in
  $\cB^d(v)$ are minima of the energy among the configurations
  with volume $v$ for all $v \le m$.
 Clearly, such an $L$ 
  exists, since the configurations winding around the torus have at
  least magnetization $L$.

 We define the set $\bar \cB^d$ of the {\it candidate saddles in
  dimension $d$} in a recursive way:
 \begin{enumerate}
 \item in one dimension it is the set of configurations consisting of
   a single plus spin in the sea of minuses.
 \item in dimension $d$ it is the set of configurations in which
   the pluses form a quasi-cube with one side of length $\ell_d-1$
   and all other sides of length $\ell_d$
   with a $(d-1)$-dimensional candidate saddle attached
   on one of the squared $(d-1)$-dimensional faces.
 \end{enumerate}
 Notice that $H(\bar \cB^d) = \max_{v \le L^d} H(\cB^d(v))$.

 Clearly, all the candidate saddles have the same volume $v^*_d$
  and are in $\cB^d(v^*_d)$.
 Moreover,  each $\eta \in \cB^d(v)$ is
  connected to a configuration in $\cB^d(v+1)$ and to one
  configuration in $\cB^d(v-1)$.

 Hence, the candidate saddles are saddles between
	  $\minus$ and $\plus$
 since from any candidate saddle there exist a path
	  leading to $\minus$ and a path leading to $\plus$
	  both reaching their maximum energy in
	  the starting point.

 The following lemma  was communicated to us by E. Olivieri [dHNOS].

 \begin{lemma}
  \label{pozzi}
  The set $\cM := \graffa{\minus,\plus}$ is a metastable set.
 \end{lemma}
 \begin{proof}
We have to show that for some $\d>0$, 
for any $\s\neq \minus$, $\G(\s)<\G(\minus)$,
i.e. for any   $\s \not \in
 \graffa{\minus,\plus}$, there exists a configuration $\s'$ such
 that
 \begin{enumerate}
 \item $H(\s') < H(\s)-\d$
 \item $\widehat H(\s,\s') - H(\s) <
	\widehat H(\minus,\plus) - H(\minus)-\d$.
 \end{enumerate}
 For $\eta \in \O$, let $|\eta|$ and $\wp(\eta)$ be the number of
 pluses and the number of pairs of nearest neighbors with different
 spin (namely, the perimeter, or cardinality of the contour),
 respectively. It is a well known fact that the Hamiltonian of the
 Ising model can be written as
 \begin{equation}
   \label{eq:h2}
   H(\eta)=\wp(\eta) - h |\eta| +H(\minus)
 \end{equation}
 Let
  $m:=\min \graffa{k \ge 1 ; \exists \eta \text{ with } |\eta|=k
  \text{ and } H(\eta) \le H(\minus)}$.
 Let
  $\o:\minus \to \plus$
  be a monotone one-dimensional subgraph such that $\o_k \in \cB^d(k)$
  that reaches its maximal energy in
  $\cS (\minus, \plus)$.
 Clearly, $H(\o_m)<H(\minus)$.
 Let $\s \cup\eta$ denote the
  configuration where
  $\tonda{\s \cup \eta} (x):= \s(x) \lor \eta(x)$
  and $\s \cap \eta$ denote the
  configuration where
  $\tonda{\s \cap \eta} (x):= \s(x) \land \eta(x)$.

 A direct computation shows that
 \begin{equation}
   \label{eq:perimetri}
   \wp(\s) + \wp(\eta) \ge \wp(\s \cup \eta) + \wp(\s \cap \eta).
 \end{equation}

 Since $\s$ is neither $\plus$ nor $\minus$, there exists at least one
 pair of nearest neighbour sites $i,j$ such that $\s(i)=-1$ and $\s(j)=+1$.
 By translation invariance we may assume that the first occupied site
 in the sequence $\o_k$ is $i$ and the second is $j$.  Thus in the
 first step,
   $\s \cap \o_1 \not = \minus$ and
 $H(\s \cup \o_1) - H(\s) < H(\o_1) - H(\minus)$, while in the second step
  $\s \cup \o_2 = \s \cup \o_1$, so that
 $|\s \cap \o_k|<k$ for all $k \ge 2$. We choose $\s'=
\s\cup\o_m$. 

 In order to prove point 2. we notice that for $k \le m$,
 \begin{eqnarray}
   H(\s \cup \o_k) - H(\s)
   &=&
   \wp(\s \cup \o_k) - \wp(\s) - h \tonda{|\s \cup \o_k| - |\s|}
   \nonumber\\
   &\le&
   \wp(\o_k) - \wp(\s \cap \o_k) - h \tonda{|\o_k| - |\s \cap \o_k|}
   \nonumber\\
   &=&
   H(\o_k) - H(\s \cap \o_k) < H(\o_k) - H(\minus),
 \label{eq:costo}
 \end{eqnarray}
 since by definition, $|\s \cap \o_k| < m$.

 For $k=m$, from \eqref{eq:costo} we get point 1. since 
 \begin{eqnarray}
   H(\s \cup \o_m) - H(\s)
   < H(\o_m) - H(\minus) \le 0
 \label{eq:scendo}
 \end{eqnarray}
 By putting together \eqref{eq:scendo} and \eqref{eq:costo},
 we see that the energy of the configuration $\s'$
 is lower than the energy of $\s$ and that the maximal energy in
 the one-dimensional subgraph $\s \cup \o_k:\s \to \s'$ is lower than
 $\widehat H(\minus,\plus)$.
 \end{proof}

 In the next Theorem, we use the results of [NS] and [BC]
 to describe the energy landscape.

 \begin{thm}
 \label{2e3}
 {\rm (from [NS] and [BC]).}
 In dimension $2$ and $3$,
 \begin{enumerate}
 \item The set of saddles between $\minus$ and $\plus$
	  coincides with the set of candidate saddles.
 \item If $2(d-1)/h$ is not an  integer, the saddles are simple.
 \end{enumerate}
 \end{thm}

 \smallskip
 \begin{conjecture}
 \label{conj1}
 {Theorem \ref{2e3} holds in any dimension}.
 \end{conjecture}

 \medskip

 We define $D_+$ as the set of all states that are
  larger  than a candidate saddle
  (namely of the form $\s \cup \eta \not = \s$
  for some candidate saddle $\s$).
 By the same procedure of the proof of Lemma \ref{pozzi}, we can
  easily see that all configurations in $D_+$ have the saddle with
  $\plus$ below the saddle with $\minus$.
 We set $\sselle:=\partial D_+$ and
 $D_-:=\O \setminus \tonda{D_+\cup \sselle}$.
 For
 any configuration in $\sselle$, we define $\px_z:= \sum_{y' \in
 D_-} P (z,y')$  and $\py_z:= \sum_{x' \in D_+} P (z,x')$. Notice
 that in a Metropolis dynamics, all the transitions associated with
 an energy gain have probability $1 / |\L|$. Hence, $\px_z$ (resp.
 $\py_z$) is exponentially close to the number of nearest neighbor
 of $z$ that are smaller (resp. larger) than $z$. Notice that
 $\sselle$ is the outer boundary of the connected
 set $D_-$ that contains $-1$. A direct computation shows that the set of 
candidate saddles coincides with the set of minima in $Z$.
 Hence, in dimensions $2$ and $3$, and whenever the Conjecture 5.4 holds,
 $\pref{\minus,\plus}$ is exponentially close to the factor
 \begin{equation}
  \label{eq:prefising}
   \sum_{z \in \cS_{\minus,\plus}}
    \frac{\px_z \py_z}{\px_z + \py_z},
 \end{equation}
 computed with respect to $\sselle$.

 Let
 \begin{equation}\label{gammad}
   \G_d:= 2\sum_{k=2}^d 
     \tonda{\tonda{k \ell_k^{k-1}-(k-1)\ell_k^{k-2}}
     -h \tonda{\ell_k^{k} - \ell_k^{k-1}}} +2 -h
 \end{equation}
 be the activation energy of the candidate saddle in dimension $d$.

 \begin{thm}
 \label{mainising} 
 For the Ising model on a (sufficiently large) $d-dimensional$ torus
 $\L(l)$,
 in dimension $d > 3$,
   there exists a constant $c_d$ such that
   \begin{equation}\label{d23i}
     \E \t_{+}^{-} = c_d e^{\b \G_d} \err
   \end{equation}
 \end{thm}
 If Conjecture \ref{conj1} holds and $sk/h$ is not an integer for all
$k=1,\dots,d-1$, 
then the pre-factor $c_d$ is equal to
 \begin{equation}
   \label{eq:cd}
   \tonda{d! \frac{2^d}{3} \tonda{1-\frac{2}{\ell_2}}
    \prod_{k=1}^{d-1}
    \tonda{\ell_{k+1} - \ell_k +1}^k }^{-1}
 \end{equation}

 \begin{lemma}
   \label{numselle}
 The number  of candidate saddles in dimension $d$
 contained in a $d$-dimensional cube
 of side-length $l \ge \ell_d$ is
 \begin{equation}
  \label{num}
   \cN_d(l) = 2^{d-1} d! \tonda{l - \ell_{d} +1}^d
   \prod_{k=1}^{d-1} \tonda{\ell_{k+1} - \ell_{k} +1}^k
 \end{equation}
 All the candidate saddles have $\px = l^{-d}$, while $\py $ can
 take the value $l^{-d}$ or $2l^{-d}$.
 The fraction  of candidate
 saddles with $\py =l^{-d}$ is $\frac{2}{\ell_2}$, independently of $d$
 and $l$.
 \end{lemma}

 \begin{proof}

 Let $\cN_d:=\cN_d(\ell_d)$.

 The key observation is that the pluses of a candidate saddle
 are contained in exactly one cube of side-length $\ell_d$.

 A $d-dimensional$ cube of side-length $l \ge \ell_d$ contains
 $(l-\ell_d+1)^d$ of such cubes.
 Hence, $\cN_d(l)=(l-\ell_d+1)^d \cN_d$.

 Given a cube of side-length $l$,
 there are $2 d$ possible choices for the incomplete face and
 $\cN_{d-1}(\ell_d)$ ways to arrange the  ($d-1$)-dimensional
 candidate droplet on this face.
 Hence,
 $
 \cN_d=2 d \cN_{d-1}(\ell_d) =
 2 d (\ell_d-\ell_{d-1}+1)^{d-1} \cN_{d-1}
 $.

 Since $\cN_1=1$, a simple calculation gives \eqref{num}.

 The computation of the number $m_d(l)$ of candidate saddles with
 $\py =l^{-d}$ is very similar: In dimension two, $m_2 (\ell_2) =
 8$ i.e. the number of configurations made of a quasi-square plus a
 protuberance at one end of one of the longest sides. All other
 candidate saddles have  $\py =2l^{-d}$, since there are two neighbors of
 the protuberance that can be occupied. All   candidate saddles
 have $\px=l^{-d}$, since we can void the occupied site and reach a
 quasi-cube in $D_-$. In general, for $d>1$, the only sites with
 $d$ plus-neighbors are in an incomplete face of the \ddim critical
 cube. Hence, $m_d(\ell_d)$ is equal the number of
 ($d-1$)-dimensional critical squares on the faces of the \ddim
 critical cube times $m_{d-1}(\ell_d)$ namely, $ m_d(\ell_d) =2 d
 (\ell_d-\ell_{d-1}+1)^{d-1} m_{d-1}(\ell_{d-1}) $. On the other
 hand, $m_d(l) =2 d (l-\ell_d+1)^d m_d(\ell_d)$. Thus, the ratio
 $m_d(l) / \cN_d(l)$ does not depend on $d$ or on $l$ and is equal
 to  $2/ \ell_2$.
 \end{proof}

 \begin{corollary}
 \label{prefattore}
 The number  of candidate saddles in dimension
 $d$ contained in a $d$-dimensional torus of side-length $l \ge
 \ell_d$ is
 \begin{equation}
  \label{num2}
   \widetilde \cN_d(l) = 2^{d-1} d! \; l^d \;
   \prod_{k=1}^{d-1} \tonda{\ell_{k+1} - \ell_{k} +1}^k
 \end{equation}
 \begin{equation}
  \label{eq:stima}
   = d! 2^{\frac{d^2+d-2}{2}} h^{-\frac{d^2-d}{2}}
     \tonda{1+o(h)}
 \end{equation}
 All the candidate saddles have $\px = l^{-d}$, while $\py $ can
 take the value $l^{-d}$ or $2l^{-d}$. The fraction  of candidate
 saddles with $\py =l^{-d}$ is $\frac{2}{\ell_2}$, independently of
 $d$ and $l$.
 \end{corollary}
 \begin{proof}
 The result is a straightforward consequence of
 lemma \ref{numselle} and of the fact that the number of
 $d$-dimensional cubes of side-length $\ell_d$
 that can be put into the torus is $l^d$.

 The estimate in \eqref{eq:stima} comes from the approximation
 $\ell_d - \ell_{d-1} + 1 = \frac{2}{h}\tonda{1+o(h)}$
 and hence \eqref{eq:stima}.
 \end{proof}

 \begin{proof}[Proof of Theorems \ref{mainising1} and \ref{mainising}:]
 The results of Theorem \ref{mainising1} are straightforward
 consequences of Theorem \ref{th1}, Lemma  \ref{2e3}, Lemma \ref{pozzi}, 
 and Corollary \ref{prefattore}. 
 In higher dimension, 
 the corresponding result comes 
 from Theorem \ref{th1}
 and Lemma \ref{pozzi}.
 If conjecture \ref{conj1} holds, Corollary \ref{prefattore} 
 gives the estimate in \ref{eq:cd}. 
 \end{proof}

 \bigskip

 In conclusion, let us notice that the form of the 
 quantities $\TT{x}{y}$ in the case of the
 Metropolis dynamics may offer an interpretation in terms
 of ``free energy of the set of saddle points''.
 Indeed, every point in $\sselle$ gives a contribution
 to the pre-factor $\pref{x,y}$ 
 that does not depend on $\b$ and can be bounded by a 
 constant $c$.
 With the arguments in the proof of Lemma \ref{lemma1},
 we get 
 \begin{equation}
   \label{eq:entropia}
   \TT{x}{y} 
    \sim
   c \cN e^{-\b \G} =
   c \exp \tonda{ -\b (\G - T \log \cN)},
 \end{equation}
 where $\cN$  is the number of saddles 
 between $x$ and $y$ and $T=\b^{-1}$.
 The logarithm of $\cN$ can be interpreted as an entropy.
 This interpretation could be related to 
 the results by Schonmann and Shlosman (see [ScSh])
 on the connections between 
 Wulff droplets and the metastable relaxation of kinetic Ising model.

 \section{\label{Appendix} Appendix.}

 In this appendix we briefly explain how our general approach can be
 generalized to situations than the saddles are more complicated
 when the isolated single points assumed in Section
 2. The point we want to make is that in such a case it is still
 possible to localize the problem to the understanding of the
 neighborhood of the saddle points and to thus reduce the analysis of
 the capacities to a `local' variational problem. Let us consider a
 situation when in the computation of a transition from $x$ to $y$ we
 encounter a set of saddles $\cS_{x,y}$ that can be decomposed into
 a collection of disconnected subsets $\cS^{(k)}$, $k=1,\dots,L$.
 By definition, it must be true that each of the sets $\cS^{(k)}$ is
 connected to two subsets $\cR^{(k)}$ and $\cN^{(k)} $ of $D^y_x$ and
 $D^x_y$, respectively.  Let us define
 \begin{equation}
 \label{app1}
 C(k):=\sum_{i\in \cN^{(k)}}e^{-\b(H(x)-H(i))}
 \wt\P\left(\t^i_{\cR^{(k)}}<\t^{i}_{\cN^{(k)}}\right)
 \end{equation}
 where $\wt \P$ is the law of the chain where all the edges exiting
 from the sets $S^{(k)}$ not leading to $\cN^{(k)}$ or $\cR^{(k)}$ are
 cut. Note that it is not difficult to see that

 \begin{equation}
 \label{app2}
 C^{(k)} = \inf_{h\in \cH^{\cR}_{\cN}}\wt\Phi(h)
 \end{equation}
 Repeating the steps of the proof of Lemma 4.2, one obtains then that

 \begin{lemma}
   \label{applem}
 In the situation described above we have that
 \begin{equation}
  \label{app3}
   \P\left(\t^x_y<\t^x_x\right)= \sum_{k=1}^L C^{(k)}\left(1+o(e^{-\b \d})\right)
 \end{equation}
 \end{lemma}

{\bf Acknowledgements:} We thank Frank den Hollander, Enzo Olivieri, 
Elisabetta Scoppola, and Raphael Cerf for useful discussions.

\end{document}